\documentclass[onecolumn,showpacs]{revtex4}

\usepackage{graphicx}
\usepackage{dcolumn}
\usepackage{bm}

\input epsf

\begin{document}

\title{\Large The study of gravitational collapse model in higher dimensional space-time}

\author{\bf Ujjal Debnath}
\email{ujjaldebnath@yahoo.com}
\author{\bf Subenoy Chakraborty}
\email{subenoyc@yahoo.co.in}
 \affiliation{Department of
Mathematics, Jadavpur University, Calcutta-32, India.}

\date{\today}

\begin{abstract}
We investigate the end state of the gravitational collapse of an
inhomogeneous dust cloud in higher dimensional space-time. The
naked singularities are shown to be developing as the final
outcome of non-marginally bound collapse. The naked singularities
are found to be gravitationally strong in the sense of Tipler .
\end{abstract}


\maketitle

The gravitation collapse is an important and challenging issue in
Einstein gravity, especially after the formation of famous
singularity theorems [1] and Cosmic Censorship Conjecture (CCC)
[2]. Also from the point of view of black hole physics and its
astrophysical implications, it is interesting to know the final
fate of gravitational collapse [3] in the background of general
relativity. The singularity theorems as such can not predict about
the visibility of the singularity to an external observer as well
as their strength. On the other hand, the CCC is incomplete [4,5]
in the sense that there is no formal proof of it as well as there
are counter examples of it. However, it has been pointed out
recently [6] that the nature of the central shell focusing
singularity depends on the choice
of the initial data.\\

The study of higher dimensional gravitational collapse (in
Tolman-Bondi form) was originated by Banerjee etal [7] and
subsequently by Ghosh and Beesham [8] but they restricted to only
self-similar and non self-similar solutions for marginally bound
case. Recently, Ghosh and Banerjee [9] have studied naked
singularity for five dimensional Tolman-Bondi model for
self-similar solution in non-marginally
bound case.\\

Moreover recently, Joshi etal [10] have shown that physically
shear responsible for the formation of naked singularity.
Subsequently Banerjee etal[11] and Debnath and Chakraborty [12]
have studied gravitational collapse in higher dimensional
Tolman-Bondi model for both marginally bound and non marginally
bound cases. They have interestingly shown that for marginally
bound case naked singularity is possible only upto five dimension
while naked singularity may be possible in all dimensions for non
marginally bound case. Then Goswami and Joshi [13] have pointed
out that this peculiar feature of naked singularity (in marginal
bound case) is due to the choice of the initial condition.\\

In the present paper we study gravitational collapse for
non-marginally bound case considering non self-similar solutions
and it is generalization to higher dimension of the Dwivedi and
Joshi [14]. The geodesic equations can not completely solved due
to the presence of the complicated hypergeometric function. So we
present numerical results which favour the formation of naked
singularity in any dimension.\\

The $n$ dimensional Tolman-Bondi metric in co-moving co-ordinates
is given by

\begin{equation}
ds^{2}=e^{\nu}dt^{2}-e^{\lambda}dr^{2}-R^{2}d\Omega^{2}_{n-2}
\end{equation}

where $\nu,\lambda,R$ are functions of the radial co-ordinate $r$
and time $t$ and $d\Omega^{2}_{n-2}$ represents the metric on the
$(n-2)$-sphere. Since we assume the matter in the form of dust,
the motion of particles will be geodesic allowing us to write
~$e^{\nu}=1$. Now from the Einstein's field equations for the
metric (1), one can obtain

\begin{equation}
e^{\lambda}=\frac{R'^{2}}{1+f(r)}
\end{equation}

and

\begin{equation}
\dot{R}^{2}=f(r)+\frac{F(r)}{R^{n-3}},
\end{equation}

where, $f(r)$ and $F(r)$ are arbitrary functions of radial
co-ordinate $r$ alone with the restriction $1+f(r)>0$ for obvious
reasons. Physically $F(r)$ and $f(r)$ represents the mass function
and the binding energy function respectively.\\

The energy density $\rho(t,r)$ is therefore given by

\begin{equation}
\rho(t,r)=\frac{(n-2)F'(r)}{2R^{n-2}R'}
\end{equation}

Since in the present discussion we are concerned with the
gravitational collapse, we require that $\dot{R}(t,r) < 0$. As it
is possible to make an arbitrary relabeling of spherical dust
shells by $r\rightarrow g(r)$ without loss of generality, we fix
the labeling by requiring that on the hypersurface $t=0$, $r$
coincides with the radius

\begin{equation}
R(0,r)=r
\end{equation}

So the initial density is given by

\begin{equation}
\rho(r)\equiv\rho(0,r)=\frac{n-2}{2}~r^{2-n}F'(r)~~\Rightarrow
F(r)=\frac{2}{n-2}\int\rho(r) r^{n-2} dr
\end{equation}

Now from eq.(4) it can be seen that the density diverges faster in
5D ($n=5$) as compared to 4D ($n=4$). For increasing dimensions
the density diverges rapidly. Hence there is relatively more
mass-energy collapsing in the higher dimensional space-time
compared to the 4D and 5D cases.\\

Integrating eq.(3) and using the relation (5), we have the
solution

\begin{equation}
t=\frac{2}{(n-1)\sqrt{F}}\left[r^{\frac{n-1}{2}}~_{2}F_{1}[\frac{1}{2},a,a+1,-\frac{f
r^{n-3}}{F}]-R^{\frac{n-1}{2}}~_{2}F_{1}[\frac{1}{2},a,a+1,-\frac{f
R^{n-3} }{F}]\right]
\end{equation}

where $_{2}F_{1}$ is the usual hypergeometric function with
$a=\frac{1}{2}+\frac{1}{n-3}$ .\\

Let us assume [14]

\begin{eqnarray}\left.\begin{array}{llll}
F(r)=r^{n-3}\lambda(r)\\\\
\alpha=\alpha(r)=\frac{r f'}{f}\\\\
\beta=\beta(r)=\frac{r F'}{F}\\\\
R(t,r)=r P(t,r)
\end{array}\right\}
\end{eqnarray}

So using equations (3), (7) and (8), we have the following
expressions

\begin{eqnarray*}
R'=\frac{1}{n-3}\left[P(\beta-\alpha)+\frac{1}{n-1}\left\{
P~_{2}F_{1}[\frac{1}{2},a,a+1,-\frac{f
P^{n-3}}{\lambda}]-P^{\frac{3-n}{2}}~_{2}F_{1}[\frac{1}{2},a,a+1,-\frac{f
}{\lambda}] \right\}\right.
\end{eqnarray*}
\vspace{-5mm}

\begin{equation}
\left.\{\alpha(n-1)-2\beta\}\sqrt{1+\frac{f P^{n-3}}{\lambda}}
+\frac{(\alpha-\beta+n-3)P^{\frac{3-n}{2}}\sqrt{1+\frac{f
P^{n-3}}{\lambda}}}{\sqrt{1+\frac{f}{\lambda}}} \right]
\end{equation}

and

\begin{eqnarray*}
\dot{R}'=\frac{1}{r}\left[\frac{1}{n-1}\left\{
~_{2}F_{1}[\frac{1}{2},a,a+1,-\frac{f
P^{n-3}}{\lambda}]-P^{\frac{1-n}{2}}~_{2}F_{1}[\frac{1}{2},a,a+1,-\frac{f
}{\lambda}] \right\}\{\alpha(n-1)-2\beta\}\right.
\end{eqnarray*}
\vspace{-5mm}

\begin{equation}
\hspace{2.2in} \left.-\alpha\sqrt{1+\frac{f P^{n-3}}{\lambda}}
+\frac{(\alpha-\beta+n-3)P^{\frac{1-n}{2}}}{\sqrt{1+\frac{f}{\lambda}}}
\right]
\end{equation}

When $\lambda(r)=$ constant and $f(r)=$ constant, the space-time
becomes self-similar. Now we restrict ourselves to functions
$f(r)$ and $\lambda(r)$ which are analytic at $r=0$, such that
$\lambda(0)>0$. From eq.(4) it can be seen that the density at the
centre $(r=0)$ is finite at any time $t$, but becomes singular at $t=0$.\\

We wish to investigate whether the singularity, when the central
shell with co-moving co-ordinate ($r=0$) collapses to the centre
at the time $t=0$, is naked .The singularity is naked if there
exists a null geodesic that emanates from the singularity. Let
$K^{a}=\frac{dx^{a}}{d\mu}$ be the tangent vector to the radial
null geodesic, where $\mu$ is the affine parameter. Then we derive
the following equations:

\begin{equation}
\frac{dK^{t}}{d\mu}+\frac{\dot{R}'}{\sqrt{1+f}}K^{r}K^{t}=0
\end{equation}

\begin{equation}
\frac{dt}{dr}=\frac{K^{t}}{K^{r}}=\frac{R'}{\sqrt{1+f(r)}}
\end{equation}

Let us define $X=\frac{t}{r}$, so that $P(t,r)=P(X,r)$. So eq.(7)
becomes

\begin{equation}
X=\frac{2}{(n-1)\sqrt{\lambda}}\left\{_{2}F_{1}[\frac{1}{2},a,a+1,-\frac{f
}{\lambda}]-P^{\frac{n-1}{2}}~_{2}F_{1}[\frac{1}{2},a,a+1,-\frac{f
P^{n-3} }{\lambda}]\right\}
\end{equation}

The nature of the singularity can be characterized by the
existence of radial null geodesics emerging from the singularity.
The singularity is at least locally naked if there exist such
geodesics and if no such geodesics exist it is a black hole. If
the singularity is naked, then there exists a real and positive
value of $X_{0}$ as a solution to the equation

\begin{eqnarray}
\begin{array}{c}
X_{0}~=\\
{}
\end{array}
\begin{array}{c}
lim~~~~~ \frac{t}{r}\\
t\rightarrow 0~ r\rightarrow 0
\end{array}
\begin{array}{c}
=~lim~~~~~ \frac{dt}{dr}\\
~~~~~~t\rightarrow 0~ r\rightarrow 0
\end{array}
\begin{array}{c}
=~lim~~~~ \frac{R'}{\sqrt{1+f}}\\
t\rightarrow 0~ r\rightarrow 0
\end{array}
\end{eqnarray}

Define $\lambda_{0}=\lambda(0), \alpha_{0}=\alpha(0), f_{0}=f(0)$,
and $Q=Q(X)=P(X,0)$. Now from equation (9) it is seen that
$\beta(0)=n-3$. We would denote $Q_{0}=Q(X_{0})$, the equations
(13) and (14) reduces to

\begin{equation}
X_{0}=\frac{2}{(n-1)\sqrt{\lambda_{0}}}\left\{_{2}F_{1}[\frac{1}{2},a,a+1,-\frac{f_{0}
}{\lambda_{0}}]-Q_{0}^{\frac{n-1}{2}}~_{2}F_{1}[\frac{1}{2},a,a+1,-\frac{f_{0}
Q_{0}^{n-3} }{\lambda_{0}}]\right\}
\end{equation}

and

\begin{eqnarray*}
X_{0}=\frac{1}{(n-3)\sqrt{1+f_{0}}}\left[\frac{1}{n-1}\left\{
Q_{0}~_{2}F_{1}[\frac{1}{2},a,a+1,-\frac{f_{0}
Q_{0}^{n-3}}{\lambda_{0}}]-Q_{0}^{\frac{3-n}{2}}~_{2}F_{1}[\frac{1}{2},a,a+1,-\frac{f_{0}
}{\lambda_{0}}] \right\}\right.
\end{eqnarray*}
\vspace{-5mm}

\begin{equation}
\left.\{\alpha_{0}(n-1)-2n+6\}\sqrt{1+\frac{f_{0}
Q_{0}^{n-3}}{\lambda_{0}}}
+\frac{\alpha_{0}Q_{0}^{\frac{3-n}{2}}\sqrt{1+\frac{f_{0}
Q_{0}^{n-3}}{\lambda_{0}}}}{\sqrt{1+\frac{f_{0}}{\lambda_{0}}}}+Q_{0}(n-3-\alpha_{0})
\right]
\end{equation}
\\

TABLE: Positive values of $X_{0}$ by eliminating $Q_{0}$ from
equations (15) and (16) for different values of the parameters
$f_{0}, \lambda_{0}$ and $\alpha_{0}$ in various dimensions.
\underline{}
\begin{center}
\begin{tabular}{lcccccc}   \hline
~$f_{0}$~~~~~~$\lambda_{0}$~~~~~$\alpha_{0}$~~~~~~~~~~~~~~~~~~~~~~~~~~~~~~~~~~~~~~Positive roots ($ X_{0}$)\\
\hline\\
~~~~~~~~~~~~~~~~~~~~~~~~4D~~~~~~~~~5D~~~~~~~~~6D~~~~~~~~~7D~~~~~~~~~8D~~~~~~~~~10D~~~~~~~~~12D~~~~~~~~~16D\\
\hline\\
-.033~~.034~~~.2~~6.75507,~~4.32936,~~3.15991,~~2.44524,~~ 1.92883,~~.669541,~~~~~~~~$-$~~~~~~~~~~~~$-$\\
~~~~~~~~~~~~~~~~~~~~~1.14476~~~1.16321~~~1.1879~~~~1.22434~~~~1.29065~~~.415056\\
\\
-.1~~~~.333~~~.2~~~3.15662,~~~1.72737,~~~~$-$~~~~~~~~~~$-$~~~~~~~~~~~$-$~~~~~~~~~~~$-$~~~~~~~~~~~~~$-$~~~~~~~~~~~~~$-$\\
~~~~~~~~~~~~~~~~~~~~~1.24003~~~~1.48602\\
\\
-.301~.302~.001~~2.07773,~~~~~~~$-$~~~~~~~~$-$~~~~~~~~~~$-$~~~~~~~~~~~$-$~~~~~~~~~~~$-$~~~~~~~~~~~~~$-$~~~~~~~~~~~~~$-$\\
~~~~~~~~~~~~~~~~~~~~~1.65625\\
\\
-.5~~~~.51~~.001~~~~~~$-$~~~~~~~~~~~~$-$~~~~~~~~$-$~~~~~~~~~~$-$~~~~~~~~~~~$-$~~~~~~~~~~~$-$~~~~~~~~~~~~~$-$~~~~~~~~~~~~~$-$\\
\\
.1~~~~~.1~~~~~2~~~~~1.667,~~~~~~1.1946~~~~~~$-$~~~~~~~~~~$-$~~~~~~~~~~~$-$~~~~~~~~~~~$-$~~~~~~~~~~~~~$-$~~~~~~~~~~~~~$-$ \\
~~~~~~~~~~~~~~~~~~~~.806578~~~~.858211\\
\\
.1~~~~~.1~~~~10~~~.463679~~~~.465426~~~~.466489~~~.466499~~~.464947~~~~.45438~~~~~~.429651~~~~~.35278\\
\\
.1~~~~~~1~~~.2~~~~~~~~~$-$~~~~~~~~~~~$-$~~~~~~~~~~$-$~~~~~~~~~~~$-$~~~~~~~~~~~$-$~~~~~~~~~~~$-$~~~~~~~~~~~~$-$~~~~~~~~~~~~$-$\\
\\
.1~~~~~~5~~~~5~~~~~~~~~$-$~~~~~~~~~~~$-$~~~~~~~~~~$-$~~~~~~~~~~~$-$~~~~~~~~~~~$-$~~~~~~~~~~~$-$~~~~~~~~~~~~$-$~~~~~~~~~~~~$-$\\
\\
~1~~~.001~~~2~~~~.996033,~~~.968869,~~~.91379,~~~.848544,~~ .783787,~~~~.669541,~~~.577489,~~~~~~~$-$\\
~~~~~~~~~~~~~~~~~~~.414324~~~~.41436~~~~~.414412~~~.414489~~~ .414605~~~~.415056~~~~.416214\\
\\
10~~.001~~.2~~~~.316014,~~~.309739,~~~~~~~~$-$~~~~~~~~~~$-$~~~~~~~~~~~$-$~~~~~~~~~~~$-$~~~~~~~~~~~~$-$~~~~~~~~~~~~$-$\\
~~~~~~~~~~~~~~~~~~~.275327~~~~.275856\\
\\
10~~~~.1~~~~5~~~~.089435~~~~.089441~~~~.089435~~~.089409~~~.08935~~~~.089061~~~~~~.088347~~~~~.084352\\
\\
10~~~~~1~~~~3~~~~.128192,~~~.128966,~~~.19718,~~~~.169063,~~.142103,~~~~~~~$-$~~~~~~~~~~~~$-$~~~~~~~~~~~~$-$\\
~~~~~~~~~~~~~~~~~~~~~~~~~~~~~~~~~~~~~~~~~~~~~~.41436~~~~~.414412~~~.414489\\
\hline\\

\end{tabular}
\end{center}

From above table we conclude the following results: \\

(i) For same value of the parameters ($f_{0}, \lambda_{0},
\alpha_{0}$) the possibility of positive real root is more in 4D
and then it decreases with increase in dimensions.\\

(ii) If $\lambda_{0}$ is positive but close to zero then it
is possible to have a naked singularity.\\

(iii) For same values of $f_{0}$ and $\lambda_{0}$, the naked
singularity in higher dimensions will less probable as we increase
the value the parameter $\alpha_{0}$.\\

The strength of singularity [15], which is the measure of its
destructive capacity, is the most important feature. A singularity
is gravitationally strong or simply strong if it destroys by
crushing or stretching any objects that fall into it. Following
Clarke and Krolak [16],we consider the null geodesics affinely
parameterized by $\mu$ and terminating at shell focusing
singularity $r=t=\mu=0$. Then it would be a strong curvature
singularity as defined by Tipler [17] if

\begin{equation}
\begin{array}{c}
lim\\
\mu\rightarrow 0\\
\end{array}
\begin{array}{c}
\mu^{2}\psi=\\
{}
\end{array}
\begin{array}{c}
lim\\
\mu\rightarrow 0\\
\end{array}
\begin{array}{c}
\mu^{2} R_{a b}K^{a}K^{b}>0\\
{}
\end{array}
\end{equation}

where $R_{a b}$ is the Ricci tensor. It is widely believed that a
space-time does not admit analytic extension through a singularity
if it is a strong curvature singularity in the above sense. Now
equation (17) can be expressed as

\begin{equation}
\begin{array}{c}
lim\\
\mu\rightarrow 0\\
\end{array}
\begin{array}{c}
\mu^{2}\psi=\\
{}
\end{array}
\begin{array}{c}
lim\\
\mu\rightarrow 0\\
\end{array}
\begin{array}{c}
\frac{(n-2)F'}{2r^{n-4} P^{n-4} R'}\left(\frac{\mu K^{t}}{R}\right)^{2}\\
{}
\end{array}
\end{equation}

Using L'Hospital's rule and using equations (9) and (10), the
equation (18) can be written as

\begin{equation}
\begin{array}{c}
lim\\
\mu\rightarrow 0\\
\end{array}
\begin{array}{c}
\mu^{2}\psi=\\
{}
\end{array}
\frac{2(n-2)(n-3)\lambda_{0}Q_{0}X_{0}(\lambda_{0}+f_{0}Q_{0}^{n-3})\sqrt{1+f_{0}}}
{\left[(n-3)\lambda_{0}X_{0}\sqrt{1+f_{0}}-(n-3)\lambda_{0}Q_{0}-
\alpha_{0}f_{0}Q_{0}^{n-2}\right]^{2}} >0\\
\end{equation}

Thus along the radial null geodesics, the strong curvature
condition is satisfied and hence it is a strong curvature singularity.\\

Finally, we conclude that formation of naked singularity will be
less probable as we increase the dimension of space-time.\\

{\bf Acknowledgement:}\\

The authors are thankful to the members of Relativity and
Cosmology Research Centre, Department of Physics, Jadavpur
University for helpful discussion. One of the  authors (U.D) is
thankful to CSIR (Govt. of India) for awarding a Junior Research Fellowship.\\

{\bf References:}\\
\\
$[1]$  S.W. Hawking and G.F.R. Ellis, The large scale structure
of space-time (Cambridge. Univ. Press,
Cambridge, England, 1973).\\
$[2]$  R. Penrose, {\it Riv. Nuovo Cimento} {\bf 1} 252 (1969);
in General Relativity, an Einstein
Centenary Volume, edited by S.W. Hawking and W. Israel (Camb. Univ. Press, Cambridge, 1979).\\
$[3]$  For recent reviews, see, e.g. P.S. Joshi, {\it Pramana}
{\bf 55} 529 (2000); C. Gundlach, {\it Living Rev. Rel.} {\bf 2}
4 (1999); A. Krolak, {\it Prog. Theo. Phys. Suppl.} {\bf 136} 45
(1999); R. Penrose, in Black holes and relativistic stars, ed. R.
M. Wald (Univ. of Chicago Press, 1998); T.P.Singh, {\it
gr-qc}/9805066; J. P. S. Lemos, {\it Phys. Lett. A} {\bf 158} 271
(1991); {\it Phys. Rev. Lett.} {\bf 68} 1447 (1992); A. IIha and
J. P. S. Lemos, {\it Phys. Rev. D} {\bf 55} 1788 (1997); A. IIha,
A. Kleber and J. P. S. Lemos, {\it J. Math.
Phys.} {\bf 40} 3509 (1999).\\
$[4]$  P.S. Joshi, Global Aspects in Gravitation and Cosmology (Oxford Univ. Press, Oxford, 1993).\\
$[5]$ C.J.S. Clarke, {\it Class. Quantum Grav.} {\bf 10} 1375
(1993);T.P. Singh, {\it J. Astrophys. Astron.}
{\bf 20} 221 (1999).\\
$[6]$  F.C. Mena, R. Tavakol and P.S. Joshi, {\it Phys. Rev. D} {\bf 62} 044001 (2000).\\
$[7]$  A. Benerjee, A. Sil and S. Chatterjee, {\it Astrophys. J.}
{\bf 422} 681 (1994); A. Sil and S. Chatterjee, {\it Gen. Rel.
Grav.} {\bf 26} 999 (1994); S. Chatterjee, A. Banerjee and B.
Bhui, {\it Phys. Lett. A} {\bf 149} 91 (1990).\\
$[8]$  S. G. Ghosh and A. Beesham, {\it Phys. Rev. D} {\bf 64}
124005 (2001); {\it Class. Quantum Grav.}
{\bf 17} 4959 (2000).\\
$[9]$  S. G. Ghosh and A. Banerjee, {\it Int. J. Mod. Phys.
D},(2002) (accepted),{\it gr-qc}/0212067 (2002).\\
$[10]$  P.S. Joshi, N. Dadhich and R. Maartens, {\it Phys. Rev. D} {\bf 65} 101501({\it R})(2002).\\
$[11]$ A. Banerjee, U. Debnath and S. Chakraborty, {\it gr-qc}/0211099 (2002) .\\
$[12]$ U. Debnath and S. Chakraborty, {\it gr-qc}/0211102 (2002) .\\
$[13]$ R. Goswami and P.S. Joshi, {\it gr-qc}/02112097  (2002) .\\
$[14]$  I. H. Dwivedi and P. S. Joshi, {\it Class. Quantum Grav.}
{\bf 9} L69 (1992).\\
$[15]$  F. J. Tipler, {\it Phys. Lett. A} {\bf 64} 8 (1987).\\
$[16]$  C. J. S. Clarke and A. Krolak, {\it J. Geom. Phys.} {\bf
2} 127 (1986).\\
$[17]$  F. J. Tipler, C. J. S. Clarke and G. F. R. Ellis, General
Relativity and Gravitation ed. A Held (New York, Plenum) 1980.\\

\end{document}